\documentclass[aps,prl,preprint]{revtex4-1}
\usepackage{graphicx}
\usepackage{amsmath, amsthm}
\usepackage{amssymb}

\begin{document}

\title{Strain induced ferroelectric topological insulator}
\author{Shi Liu}
\affiliation{The Makineni Theoretical Laboratories, Department of Chemistry, University of Pennsylvania, Philadelphia, PA 19104--6323 USA}

\author{Youngkuk Kim}
\affiliation{The Makineni Theoretical Laboratories, Department of Chemistry, University of Pennsylvania, Philadelphia, PA 19104--6323 USA}

\author{ Liang Z. Tan}
\affiliation{The Makineni Theoretical Laboratories, Department of Chemistry, University of Pennsylvania, Philadelphia, PA 19104--6323 USA}

\author{Andrew M. Rappe}
\email{rappe@sas.upenn.edu}
\affiliation{The Makineni Theoretical Laboratories, Department of Chemistry, University of Pennsylvania, Philadelphia, PA 19104--6323 USA}

\begin{abstract}
 The simultaneous presence of seemingly incompatible properties of solids often provides a unique opportunity to address questions of fundamental and practical importance.  The coexistence of ferroelectric and topological orders is one such example. Ferroelectrics, which have a spontaneous macroscopic polarization switchable by an applied electric field, usually are semiconductors with a well-developed wide band gap~\cite{Piskunov04p165,Bennett08p17409,Wang14p152903} with a few exceptions~\cite{Grinberg13p509,Bennett12p167602,Xu11p37006}. On the other hand, time-reversal symmetric $Z_2$ topological insulators (TI)~\cite{Hasan10p3045}, characterized by robust metallic surface states protected by the topology of the bulk, usually are narrow-gap semiconductors ($< 0.7$ eV)~\cite{Ando13p102001,Yan13p709} which allow band inversion induced by the spin-orbit interaction. To date, a ferroelectric topological insulator (FETI) has remained elusive, owing to the seemingly contradictory characters of the ferroelectric and topological orders. Here, we report that the FETI can be realized in halide perovskite CsPbI$_3$ under strain. Our first-principles study reveals that a non-centrosymmetric ferroelectric structure of CsPbI$_3$ is energetically favored under a wide range of pressures, while maintaining its topological order. The proposed FETI is characterized by switchable polar surfaces with spin-momentum locked Dirac cones, which allows for electric-field control of topological surface states (TSSs) and the surface spin current. Our demonstration of a FETI in a feasible material opens doors for future studies combining ferroelectric and topological orders, and offers a new paradigm for diverse applications in electronics, spintronics, and quantum information. 
 \end{abstract}

\maketitle 
A TI is a topological state of matter that is characterized by metallic boundary modes (edge or surface) arising from the non-trivial topology of an insulating bulk electronic structure~\cite{Hasan10p3045}. The metallic nature of the boundary states is protected by time-reversal symmetry, providing transport channels that are robust against ordinary non-magnetic impurities. Fundamental scientific interest and potential applications of TIs in electronics, spintronics, and quantum information have driven tremendous efforts to find robust material realizations of TIs. To date, more than 50 candidate materials~\cite{Zhu12p235401} have been suggested theoretically in diverse material classes with some confirmed experimentally,  including chalcogenides~\cite{Fu07p106803, Hsieh08p970, Zhang09p438, Chen09p178, Xi09p398}, oxides~\cite{Yan13p709,Shitade09p256403, Carter12p115105, Yang10p085111}, ternary Heusler compounds~\cite{Chadov10p541}, halide perovskites~\cite{Yang12p614,Jin12p121102}, and a non-centrosymmetric Rashba semiconductor~\cite{Bahramy12p679}. A characteristic feature common to these materials is a narrow band gap originating from the band inversion mechanism. Without the spin-orbit interaction, these materials would be normal insulators (NIs); the spin-orbit interaction interchanges the characters of the valence and conduction bands, resulting in a non-trivial topology. For this to occur, the band gap must be smaller than the strength of the spin-orbit interaction, typically less than 1 eV, in order to host TIs.  On the other hand, ferroelectricity is usually found in transition metal oxides with wide band gaps, due to the large difference in electronegativity between oxygen and transition metals. Most technologically important ferroelectric materials (e.g., PbTiO$_3$ and BaTiO$_3$) have band gaps larger than 3 eV~\cite{Bennett08p17409}. The lowest band gap for a ferroelectric oxide is 1.39~eV, obtained recently in the single-phase solid oxide solution [KNbO$_3$]$_{1-x}$[BaNi$_{1/2}$Nb$_{1/2}$O$_{3-\delta}$]$_x$(KBNNO)~\cite{Grinberg13p509}. 

While the observed narrow band gap in various TIs seems to exclude the co-existence of ferroelectricity, there is no fundamental physics preventing the realization of both ferroelectric and topological order in a single-phase material. A ferroelectric topological insulator that combines switchable polarization and robust TSSs 
is likely to support novel device functionalities. Here, we demonstrate through first-principles calculations that the FETI can be realized in cubic halide perovskite CsPbI$_3$ under strain in the absence of PbI$_6$ octahedron rotation. The consequences of the coexistence of ferroelectricity and the non-trivial topology are explored. We find that the TSSs in FETI are tunable by the electrostatic potential associated with the polarity of the surfaces, which allows for simplified nanoscale engineering of topological {\em p}-{\em n} junctions~\cite{Wang12p235131} when combined with domain engineering.~\cite{Xu15p79} 

Using density functional theory (DFT) calculations (see Methods), we find that the cubic halide perovskite CsPbI$_3$ experiences a paraelectric--ferroelectric structural phase transition under strain. In the absence of PbI$_6$ octahedron rotation, the cubic CsPbI$_3$ with ferroelectric distortions (Pb atom displacing away from the center of PbI$_6$ octahedron) has lower formation energy ($\approx$ 10~meV per unit cell) than the paraelectric CsPbI$_3$ for a wide range of lattice constants ($a < 6.05$~\AA ) at 0~K. Free energy calculations including zero-point energy and the entropy contribution from phonons at room temperature also indicate that the system should undergo a ferroelectric distortion when the lattice constant $a$ is compressed smaller than $\approx5.90$~\AA. 

Pressure also induces a topological phase transition from a normal insulator to a topological insulator in CsPbI$_3$. Yang et al.\ and Jin et al.\ predicted and have shown that paraelectric CsPbI$_3$ is a topological insulator under hydrostatic pressure \cite{Yang12p614,Jin12p121102}. We find that the electronic structure of ferroelectric CsPbI$_3$ at $a < 6.04 $~\AA\ is adiabatically connected to the topological phase of the paraelectric CsPbI$_3$. More specifically, as shown in Fig.~1c, the band gap remains open during a structural deformation from the paraelectric topological phase to the ferroelectric phase at $a=5.79$~\AA. The bulk band structures of ferroelectric CsPbI$_3$, calculated with varying $a$, also show evidence of the topological phase transition via gap closure. 
The bulk gap closes at $a = 6.04$~\AA (Fig.~1b) , exhibiting a nodal point along the $R$--$X$ line, and reopens at $a < 6.04$~\AA. Interestingly, a nodal point occurs not only along the high-symmetric direction, but also along an arbitrary direction away from $R$ on the $k_z = \pi/a$ plane, forming a line node encircling $R$. As previously discussed \cite{Murakami07p356}, it is a characteristic feature of topological phase transition between NI and TI in three-dimensional non-centrosymmetric systems that the gapless states, in general, appear off the time-reversal invariant point ($R$) in one-dimensional parameter subspace of the ($\mathbf{k}, m$) space, where m is an external parameter tuning a topological phase transition. Whereas the specific geometry of nodal points can depend on the details of crystalline symmetries, their formation again supports a ferroelectric topological phase at a < 6.04 \AA. We directly calculate the $Z_2$ topological invariants \cite{Fu07p106803}, employing the method proposed in Ref.~\citenum{Soluyanov11p235401}. The indices are $(\mu_0;\mu_1\mu_2\mu_3) = (1;111) $, further confirming that ferroelectric CsPbI$_3$ is a strong topological insulator under high pressure. As band gaps are systematically underestimated within DFT, electronic correlations introduced by higher level theories such as the GW approximation are expected to increase the band gap and hence lower the critical value of $a$ at which CsPbI$_3$ becomes a TI\cite{Jin12p121102}. Since strain favors the ferroelectric phase (in the absence of PbI$_6$ cage rotation), CsPbI$_3$ is likely to be a FETI even in the presence of electron-electron correlations. 

To illustrate the TSSs of a FETI, we have calculated the surface band structures using a slab model (see Methods). Ferroelectric materials possess three different types of surfaces depending on the orientation of the polarization: non-polar $P^0$, positive-polar $P^+$ and negative-polar $P^-$ surfaces. We first investigate non-polar surface states by exposing the (001) surfaces with in-plane [100] polarization (Fig.~2a). As shown in Fig.~2b, the surface states exhibit a Dirac-cone-like energy dispersion at the $\overline{\rm M}$ point (0.5,0.5) of the surface Brillouin zone, confirming the nature of TSSs. 

The surface states of the polar surfaces are studied using a (100) slab with  [100]-oriented polarization (Fig.~3a). Figure 3b shows the surface band structure, with states from $P^+$ (red) and $P^-$ (blue) surfaces resolved separately for bands close to the Fermi energy (see Methods). The out-of-plane polarization breaks the degeneracy of the surface states, shifting upward in energy the Dirac cone ($p$-type) associated with the $P^-$ surface and downward in energy the Dirac cone ($n$-type) associated with the $P^+$ surface. This feature resembles the electronic structures found in graphene/ferroelectric/graphene composite~\cite{Baeumer15p6136}, and gives rise to nontrivial spin textures for bands near the Fermi energy. For instance, the surface states of the $P^-$ and ${P^+}$ surfaces are expected to have the same spin helicity at the Fermi energy, as shown schematically in the inset of Fig.3b. This is confirmed by our first-principles calculations (Fig.~3c). Similar to non-ferroelectric topological insulators, currents on the $P^-$ and $P^+$ surfaces carry opposite spins (Fig.~3d). Ferroelectricity allows these currents to be separated in energy as well. 
These results demonstrate that strained CsPbI$_3$ is a FETI with TSSs strongly coupled to bulk polarization.

Compared to a ferroelectric normal insulator (FENI), the FETI is expected to have more robust ferroelectricity. As shown in Fig.~4a, normal ferroelectrics (e.g., PbTiO$_3$) are unstable under ideal open-circuit conditions (in vacuum) due to the large depolarization field. In practice, normal ferroelectrics are stabilized by electrodes or surface charges which eliminate the effect of the bound charges. An advantage of the FETI over the FENI is the guaranteed presence of metallic surface states, which can serve as innate metallic electrodes to create an intrinsic short-circuit condition, allowing for automatic surface charge dissipation and hence stable ferroelectricity in vacuum.

The polarity of ferroelectric surfaces can be tuned and reversed by well-controlled external stimuli, such as electric field, stress field and temperature. Therefore, the coupling of switchable polarization and TSSs in FETI offers a powerful paradigm for novel device engineering. For example, the 180$^{\circ}$ domain wall in FETI may serve as a Veselago lens~\cite{Veselago68p509} for precise focusing of electric current. Shown in Fig.~4b, the 180$^{\circ}$ domain wall separates domains with opposite polarities, thus creating an $n$-type Dirac cone on the $P^+$ surface and a $p$-type Dirac cone on the $P^-$ surface. Perfect focusing of electric current occurs when the $n$-type and $p$-type Fermi circles have the same radius~\cite{Cheianov07p5816}. While this condition can only be obtained by careful tuning of local chemical potentials in graphene, it is automatically satisfied in FETIs because of the antisymmetric nature of the $P^+$  and $P^-$ surfaces. By moving the position of the domain wall via a gate voltage, it is possible to controllably change the focusing point on the $P^-$ surface. 

In addition, ferroelectric materials can be patterned into designed multi-domain configurations. This enables simplified fabrication and control of high-density {\em p}-{\em n} junction arrays at the nanoscale (Fig. 4c), which may find applications in electronics and spintronics, such as electron beam supercollimation~\cite{Park08p2920,Park08p213}. The operating bandwidth of Dirac fermion-based electron beam supercollimators increases as the period of the {\em p}-{\em n} arrays decreases~\cite{Park08p2920}. This has proved to be a limiting constraint in designing graphene-based electron beam supercollimators using conventional metallic gates which are usually larger than 100 nm in size. In contrast, nanoscale ferroelectric domains can be easily created and manipulated~\cite{Catalan12p119,Xu15p79}, opening possibilities for spin-sensitive electronic devices of wide operating bandwidth.

In conclusion, our {\em ab initio} calculations demonstrate that it is possible to realize ferroelectricity and robust topological surface states in a single material.  The electronic structure of cubic CsPbI$_3$ with ferroelectric distortions at high pressure can act as a prototype for designing ferroelectric topological insulators that are stable at ambient conditions. Replacing Cs with larger cations (e.g., molecular cation, (CH$_3$)$_4$N$^+$) may help to suppress PbI$_6$ octahedron rotation. The discovery of robust material realization of ferroelectric topological insulators can potentially lead to new device applications in electronics and spintronics.

\section{Methods}
All DFT calculations are carried out with plane-wave density functional theory package QUANTUM--ESPRESSO~\cite{Giannozzi09p395502} with Perdew-Burke-Ernzerhof  (PBE)~\cite{Perdew96p3865} density functional and norm conserving, optimized~\cite{Rappe90p1227}, designed nonlocal~\cite{Ramer99p12471} pseudopotentials generated from the OPIUM package. A plane-wave energy cutoff of 50 Ry is used. A five-atom unit cell is used for constructing the structural and topological phase diagram of cubic CsPbI$_3$ with the PbI$_6$ octahedron rotation suppressed. An $8\times8\times8$ Monkhorst-Pack $k$-point grid is used to converge the charge density, while an $8\times8\times8$ Monkhorst-Pack coarse grid, interpolated to a $32\times32\times32$ fine grid, is used for calculations of the phonon entropy. A supercell slab method is used to calculate the band structures of surface states. Atoms in the slab model are fixed to their bulk atomic positions. The supercell for modeling the non-polar surface consists of 30 layers of ferroelectric CsPbI$_3$ and 22~\AA~of vacuum. The slab has surface normal along the [001] direction and the polarization along the [100] direction. The polar surfaces are modeled with a slab that has both surface normal and polarization oriented along the [100] direction. The vacuum inside the supercell is 15~\AA~thick, and the dipole correction~\cite{Bengtsson99p12301} is applied to remove the spurious interaction between different images. The surface band structure of polar surfaces is projected onto orthogonalized atomic wavefunctions, and the weight of atomic wavefunctions from $P^+$ and $P^-$ surface atoms are then evaluated separately for 30 highest valence bands and 10 lowest conduction bands. 

\section{Acknowledgements}
S.L. was supported by the NSF through Grant CBET-1159736. 
Y.K. was supported by the NSF through Grant DMR-1120901. 
L.Z.T. was supported by the US DOE under Grant DE-FG02-07ER15920.
A.M.R. was supported by the US ONR under Grant N00014-14-1-0761. Computational support was provided by the US DOD through a Challenge Grant from the HPCMO, and by the US DOE through computer time at NERSC. 

%

\begin{figure}[p!]
   \includegraphics[width=1\textwidth]{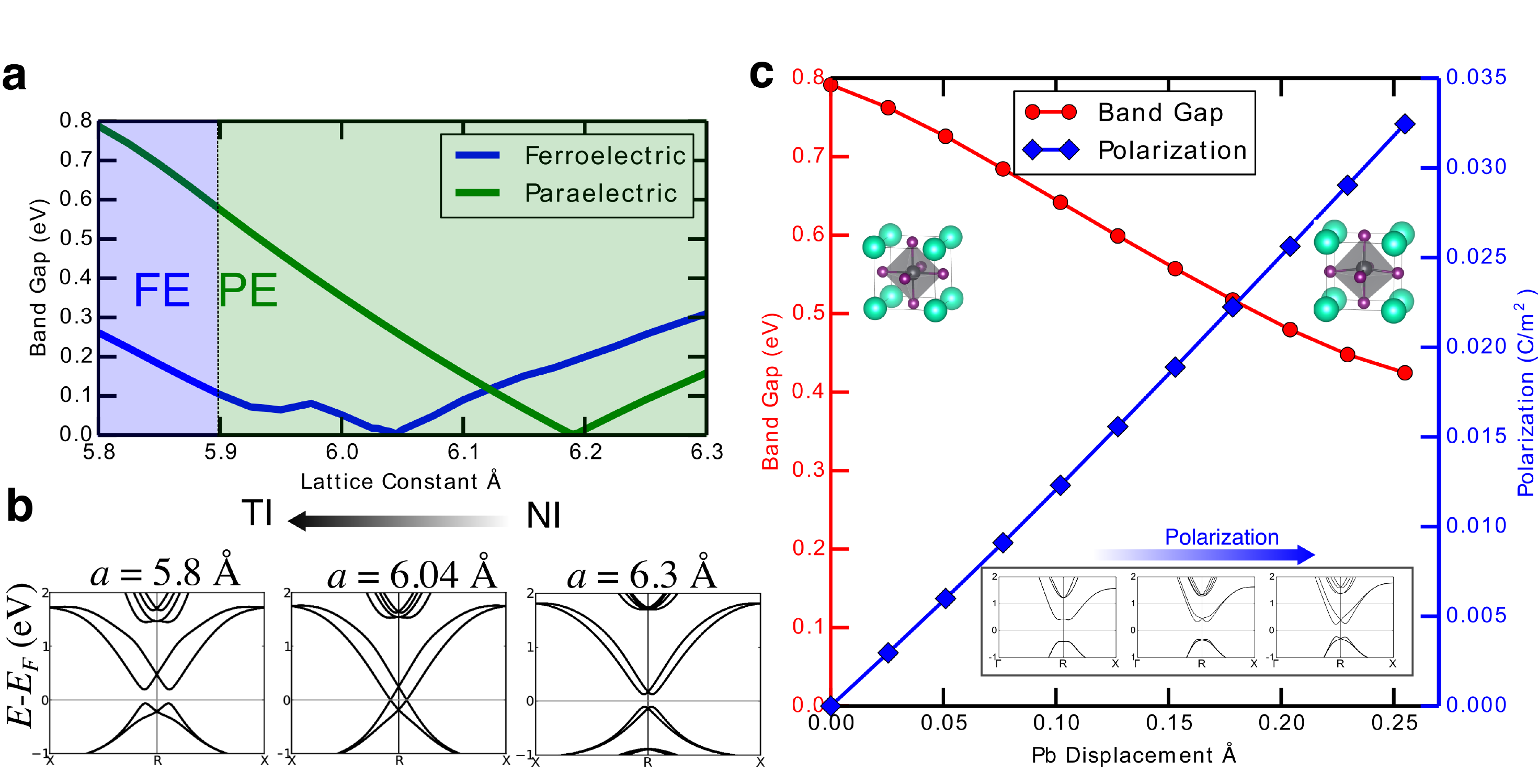}
      \caption{{\bf $|$ Pressure induced parelelectric--ferroelectric and trivial--non-trivial phase transitions in CsPbI$_3$.} {\bf a}, Electronic band gaps of paraelectric and ferroelectric CsPbI$_3$ as a function of lattice constant. Paraelectric CsPbI$_3$ has band gap closure at $a=6.19$~\AA. Ferroelectric CsPbI$_3$ changes from a normal insulator (NI) to a topological insulator (TI) at $a=6.04$~\AA. The system undergoes a paraelectric to ferroelectric structural phase transition as the lattice constant becomes smaller than $\approx5.90$~\AA~at room temperature. {\bf b}, Evolution of band structure of ferroelectric CsPbI$_3$ as a function of lattice constant. At $a=6.04$~\AA,  a line node (not shown) appear near the time-reversal invariant point ($R$) in the Brillouin zone. {\bf c}, Electronic band gap and polarization for interpolated structures connecting paraelectric and ferroelectric CsPbI$_3$ with lattice constants fixed at 5.79~\AA. Along the path, structures have gradually increasing polarization and decreasing band gaps. The inset shows the adiabatic evolution of the band structure from paraelectric to ferroelectric phase. The manifold of valence bands can be continuously transformed from paraelectric to ferroelectric without closing the band gap. }
  \label{fig1}
   \end{figure}

\begin{figure}[p!]
   \includegraphics[width=1\textwidth]{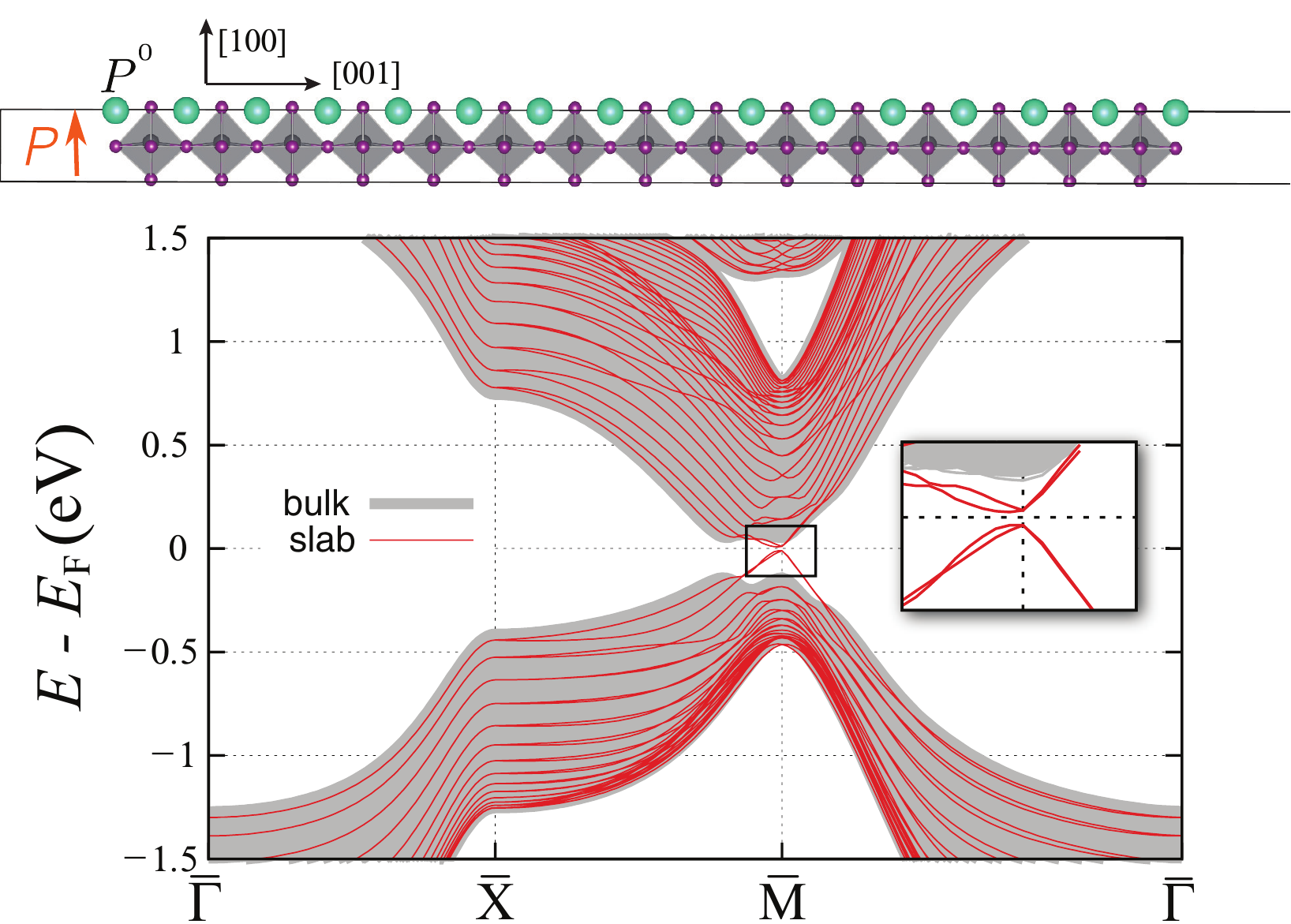}
      \caption{{\bf $|$ Electronic structure of non-polar surfaces of CsPbI$_3$.} The surface normal is along the [001] direction. The shaded (grey) bands correspond to bulk bands. The red lines highlight the slab bands. The topological surface states emerge inside the bulk band gap. The tiny gap of the surface states  (magnified in inset) appears due to the finite size of slab thickness}
  \label{fig2}
   \end{figure}
   
\begin{figure}[p!]
   \includegraphics[width=1\textwidth]{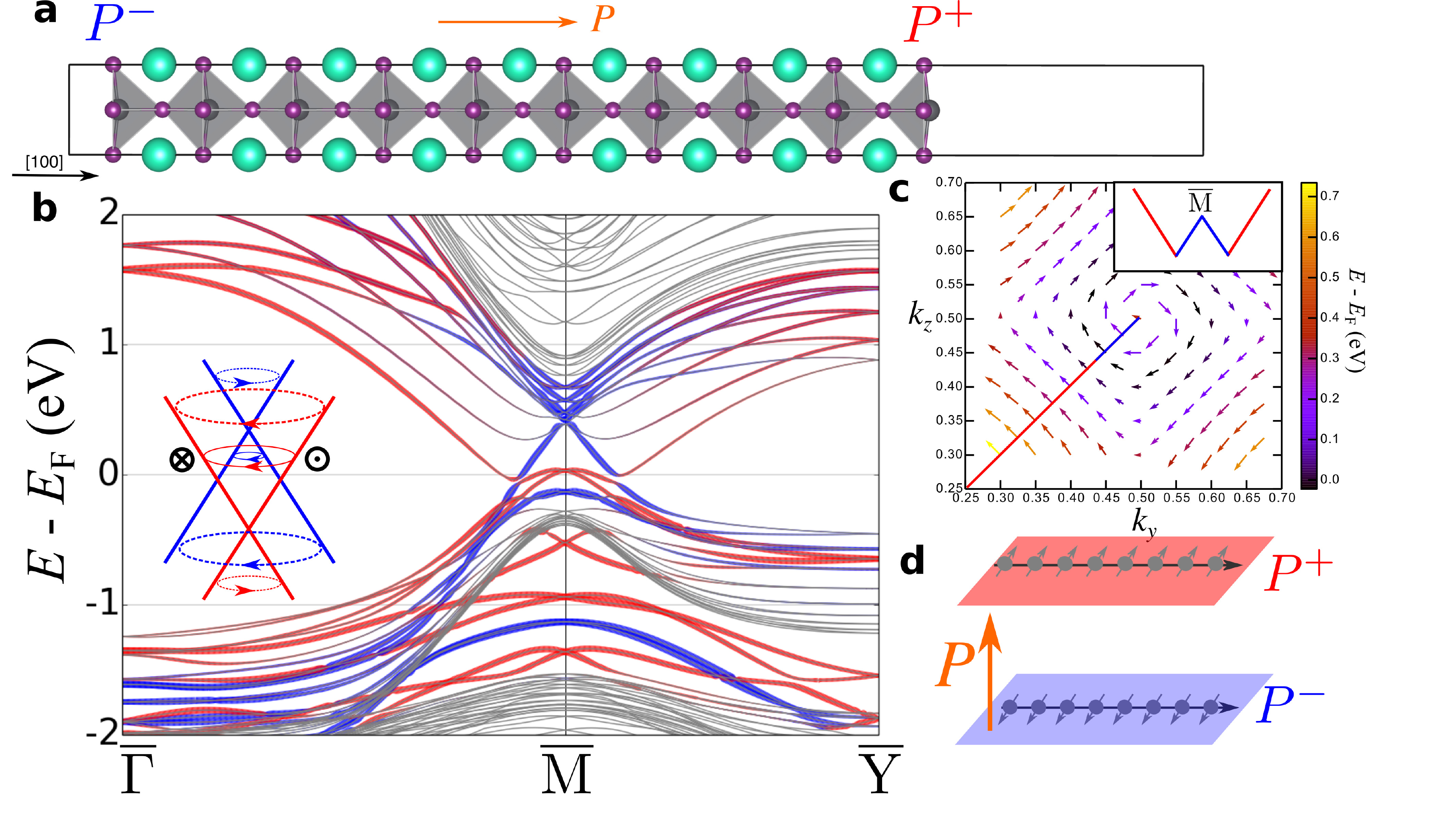}
      \caption{{\bf $|$ Electronic structure of polar surfaces of CsPbI$_3$.}  {\bf a}, Schematic of the slab geometry with surface normal and polarization along the [100] direction, creating $P^-$ surface (left) and $P^+$ surface (right). {\bf b}, Surface band structure. The red and blue lines represent the contributions of surface states from $P^+$ surface and $P^-$ surface, respectively. The thickness of lines scales with the contribution. The inset illustrates the vertically shifted Dirac cones on the $P^+$ and $P^-$ surfaces. {\bf c}, DFT spin texture for the lowest conduction band around $\overline{\rm M}$. The solid line denotes the $\overline{\Gamma}$-$\overline{\rm M}$ direction. {\bf d}, Orientations of spin currents on polar surfaces. The spins on $P^+$ and $P^-$ surfaces adopt opposite orientations for spin currents along the same direction. }
  \label{fig3}
   \end{figure}
   
   \begin{figure}[p!]
   \includegraphics[width=1\textwidth]{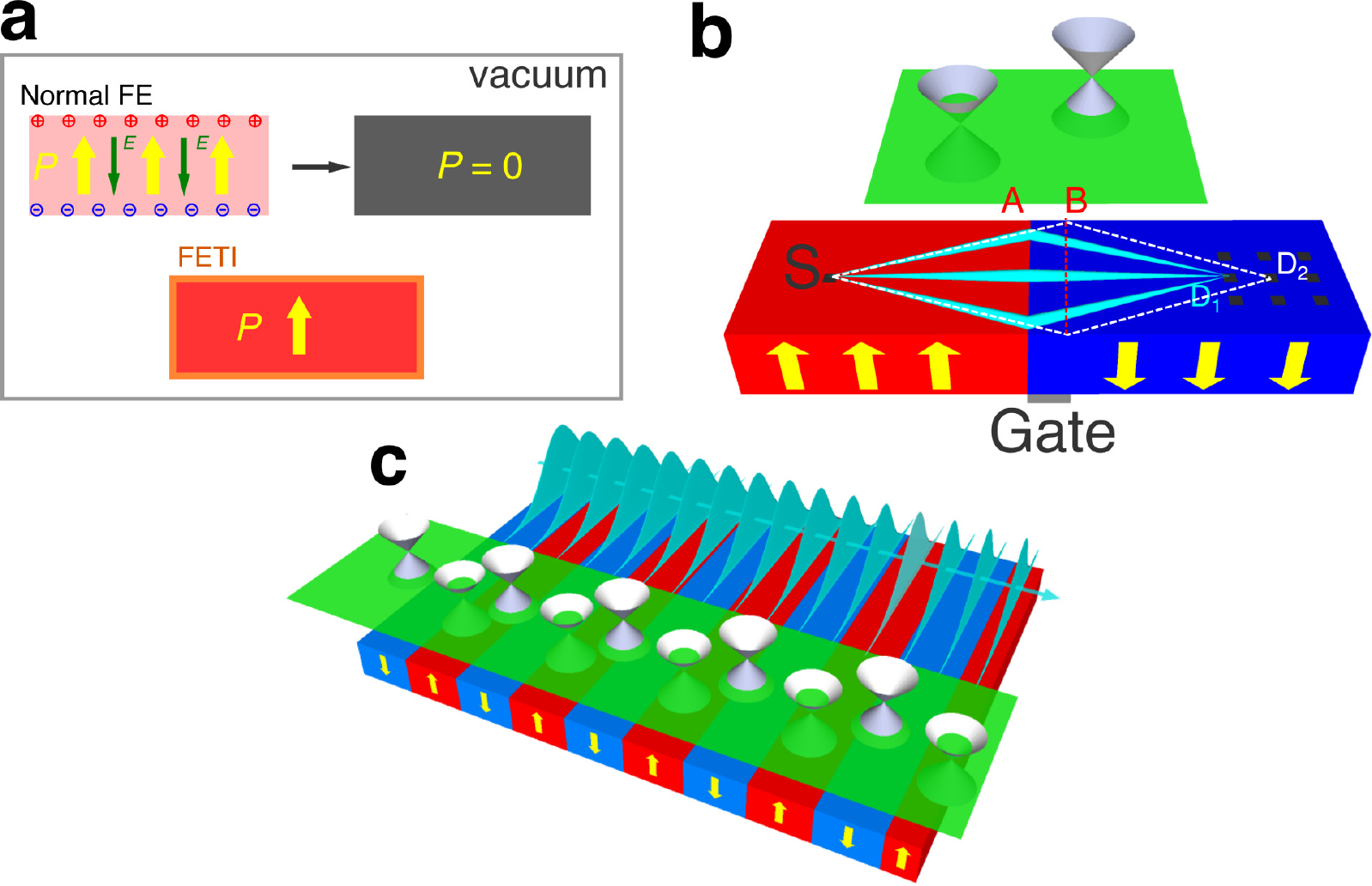}
      \caption{{\bf $|$ Schematic of functional devices based on ferroelectric topological insulator.}  {\bf a}, Robust ferroelectricity of FETI under open circuit condition. Ferroelectric thin film normal insulator (FENI) becomes paraelectric in vacuum due to the large depolarization field. The polarization of FETI is protected with metallic surface states serving as intrinsic metallic leads. {\bf b}, Domain wall in FETI as Veselago lens. Electrons injected at {\bf S} will focus at {\bf D$_1$} for domain wall positioned at line A. The domain wall can be moved to line B by reversing the direction of the polarization through a gate voltage. The focusing point will consequently be shifted to {\bf D$_2$}. {\bf c}, Ferroelectric domain arrays as nanoscale topological {\em p}-{\em n} junction arrays for electron beam supercollimation. The potential periodicity and amplitude can be tuned by varying the size of domain and the magnitude of the polarization respectively.}
  \label{fig4}
   \end{figure}

\end{document}